\documentclass{amsart}

\usepackage{graphicx}
\usepackage{flafter}

\newtheorem{theorem}{Theorem}[section]
\newtheorem{lemma}[theorem]{Lemma}
\newtheorem{corollary}[theorem]{Corollary}
\newtheorem{proposition}[theorem]{Proposition}

\theoremstyle{definition}
\newtheorem{definition}[theorem]{Definition}
\newtheorem{example}[theorem]{Example}

\newtheorem{claim}{Claim}

\theoremstyle{remark}
\newtheorem{remark}[theorem]{Remark}

\numberwithin{equation}{section}

\newcommand{\vect}[1]{\mathbf{#1}}
\newcommand{\mathtiny}[1]{\mbox{\tiny${#1}$}}

\newcommand{\eexample}{\hfill $\triangle$}
\newcommand{\del}{\partial}

\newcommand{\vc}{\vect{c}}
\newcommand{\vp}{\vect{p}}

\newcommand{\vx}{\vect{x}}
\newcommand{\vy}{\vect{y}}

\newcommand{\vzero}{\vect{0}}
\newcommand{\vomega}{\mbox{\boldmath$\omega$}}
\newcommand{\vnu}{\mbox{\boldmath$\nu$}}

\newcommand{\tva}{\tilde{\vect{a}}}
\newcommand{\tvc}{\tilde{\vect{c}}}

\newcommand{\hvc}{\hat{\vect{c}}}
\newcommand{\hvy}{\hat{\vect{y}}}

\newcommand{\hC}{\widehat C}
\newcommand{\hT}{\widehat T}

\newcommand{\hzeta}{\hat{\zeta}}

\newcommand{\F}{\mathbb{F}}
\newcommand{\N}{\mathbb{N}}
\newcommand{\R}{\mathbb{R}}
\newcommand{\Z}{\mathbb{Z}}

\newcommand{\cK}{\mathcal{K}}

\newcommand{\tC}{\widetilde{C}}
\newcommand{\tE}{\widetilde{E}}
\newcommand{\tF}{\widetilde{F}}
\newcommand{\tG}{\widetilde{G}}
\newcommand{\tH}{\widetilde{H}}
\newcommand{\tT}{\widetilde{T}}
\newcommand{\tV}{\widetilde{V}}
\newcommand{\tX}{\widetilde{X}}

\newcommand{\te}{\tilde{e}}
\newcommand{\tf}{\tilde{f}}
\newcommand{\tv}{\tilde{v}}
\newcommand{\tx}{\tilde{x}}

\newcommand{\tr}{\mathsf{T}}

\begin{document}

\title[Characterizations of pseudo-codewords of LDPC codes]
      {Characterizations of pseudo-codewords \\ of LDPC codes} 

\author[R.\ Koetter]{Ralf Koetter}
\address{Coordinated Science Laboratory, University of Illinois,
  Urbana, IL, 61801 } 
\email{koetter@uiuc.edu}
\thanks{The first author was supported in part by NSF Grant \#CCR-9984515.}

\author[W.-C.\ W.\ Li]{Wen-Ching W.\ Li}
\address{Department of Mathematics, Pennsylvania State University,
  University Park, PA 16802-6401}
\email{wli@math.psu.edu}
\thanks{The second author was supported in part by NSA Grant
  \#MDA904-03-1-0069 and NSF Grant \#DMS-0457574.} 

\author[P.\ O.\ Vontobel]{Pascal O.\ Vontobel}
\address{Department of Electrical and Computer Engineering, University
  of Wisconsin,
  Madison, WI 53706}
\email{vontobel@ece.wisc.edu}
\thanks{The third author was supported in part by NSF Grants \#CCR
  99-84515, \#CCR 01-05719, and \#ATM-0296033 and by DOE SciDAC and
  ONR Grant \#N00014-00-1-0966.}  

\author[J.\ L.\ Walker]{Judy L.\ Walker}
\address{Department of Mathematics, University of Nebraska,
  Lincoln, NE 68588-0130}
\email{jwalker@math.unl.edu}
\thanks{The fourth author was supported in part by NSF Grant
  \#DMS-0302024.} 


\date{\today}


\begin{abstract} An important property of high-performance, low
  complexity codes is the existence of highly efficient algorithms for
  their decoding. Many of the most efficient, recent graph-based
  algorithms, e.g.\ message passing algorithms and decoding based on
  linear programming, crucially depend on the efficient representation
  of a code in a graphical model.  In order to understand the
  performance of these algorithms, we argue for the characterization
  of codes in terms of a so called {\em fundamental cone} in Euclidean
  space which is a function of a given parity check matrix of a code,
  rather than of the code itself.  We give a number of properties of
  this fundamental cone derived from its connection to unramified
  covers of the graphical models on which the decoding algorithms
  operate. For the class of cycle codes, these developments naturally
  lead to a characterization of the fundamental polytope as the
  Newton polytope of the Hashimoto edge zeta function of the
  underlying graph.
\end{abstract}

\maketitle

\section{Introduction and Background}\label{section:overview}

Whenever information is transmitted across a channel, we have to
ensure its integrity against errors. While data may originate in a
multitude of applications, at some core level of the communication
system, it is usually encoded as a string of zeros and ones of fixed
length. Protection against transmission errors is provided by
intelligently adding redundant bits to the information symbols, thus
effectively restricting the set of possibly transmitted sequences of
bits to a fraction of all possible sequences. The set of all possibly
transmitted data vectors is called a code, and the elements are called
codewords.  A classical measure of goodness of a code is the code's
{\em minimum Hamming distance}, i.e., the minimum number of
coordinates in which any two distinct codewords differ. In fact, a
large part of traditional coding theory is concerned with finding the
fundamental trade-offs between three parameters: the length of the
code, the number of codewords in the code, and the minimum distance of
the code.

It is well-known that the minimum Hamming distance $d$ of a code
reflects its guaranteed error-correcting capability in the sense that
any error pattern of weight at most $\lfloor\frac{d-1}{2}\rfloor$ can
be corrected.  However, most codes can, with high probability, correct
error patterns of substantially higher weight. This insight is the
cornerstone of modern coding theory which attempts to capitalize on
the full correction capability of a code. One of the most successful
realizations of this phenomenon is found in binary low-density
parity-check (LDPC) codes. These codes come equipped with an iterative
message-passing algorithm to be performed at the receiver's end which
is extremely efficient and corrects, with high probability, many more
error patterns than guaranteed by the minimum distance.

In this situation, we are left with the problem of finding new
mathematically precise concepts that can take over the role of minimum
Hamming distance for such high performance codes. One of the main
contributions of this paper is the identification of such a concept,
namely, the {\em fundamental cone} \cite{Koetter:Vontobel} of a
code. Interestingly, the same cone appears when one is considering
low-complexity decoding approaches based on solving relaxations of
linear programs for maximum-likelihood decoding
\cite{Feldman:Wainwright:Karger:05:1}. We give here a brief motivation
of the concept.

As a binary linear code, an LDPC code $C$ is defined by a parity-check
matrix $H$. The strength of the iterative decoding algorithm, i.e.,
its low complexity, comes from the fact that the algorithm operates
locally on a so-called Tanner graph representing the matrix
$H$. However, this same fact also leads to a fundamental weakness of
the algorithm: because it acts locally, the algorithm cannot
distinguish if it is acting on the graph itself or on some finite
unramified cover of the graph. This leads to the notion of {\em
pseudo-codewords}, which arise from codewords in codes corresponding
to the covers and which compromise the decoder. Thus to understand the
performance of LDPC codes, we must understand the graph covers and the
codes corresponding to them. As will be seen later in the paper, this
is tantamount to understanding a cone in $\R^n$ defined by
inequalities arising from $H$, called the {\em fundamental cone}.  We
show that the pseudo-codewords of $C$ (with respect to $H$ and the
associated Tanner graph) are precisely the integral points in the cone
which, modulo 2, reduce to the codewords of $C$.

We emphasize below a few properties of the fundamental cone which
appear to be central to a crisp mathematical characterization.  A
recurring theme is that these properties depend upon the
representation of the code as the kernel of a given parity-check
matrix, and not solely upon the code itself as a vector space.  This
showcases the modern viewpoint of coding theory: whereas, classically,
the quality of a code was measured in terms of properties (e.g.,
length, dimension, minimum distance) of the collection of codewords
comprising the code, the quality of a code is now measured in terms of
properties (e.g., existence of pseudo-codewords of small weight) of a
particular representation of the code.  Thus, from the modern,
algorithmic point of view, a given collection of codewords might be
described by two different parity check matrices, one of which might
be considered to be very good while another would be very bad.

\begin{itemize}
\item The fundamental cone depends on the representation chosen for
  the code in terms of a parity-check matrix.  Note that a linear code
  has many different parity-check matrices and hence many different
  cones. This reflects the property of message-passing algorithms that
  both the complexity and the performance are functions of the
  structure and, in particular, the sparsity of the parity-check
  matrix.

\item The fundamental cone is an essentially geometric concept
  relating only to the parity-check matrix and independent of the
  channel on which the code is employed. Thus we can study codes and
  their parity-check matrices independently of a specific application.

\item The fundamental cone has close ties with well-established
  mathematical objects. If the parity-check matrix is chosen to be the
  (highly redundant) matrix containing all words in the dual of the
  given code, it is readily identified as the metric cone of a binary
  matroid \cite[ch.27]{Deza_Laurent}, and it is well-studied in this
  special case. Furthermore, for the particular class of LDPC codes
  called {\em cycle codes}, it is shown in
  \cite{Koetter:Li:Vontobel:Walker:1} that the fundamental cone is 
  identified with the Newton polyhedron of Hashimoto's {\em edge zeta
  function} \cite{Hashimoto} of the {\em normal graph} associated to
  the Tanner graph of the code.
\end{itemize}

The last bullet above implies that the pseudo-codewords of a cycle
code can be read off from the monomials occurring in the power series
expansion of the associated zeta function. This gives another
characterization of the pseudo-codewords for cycle codes. Inspired by
this result, we draw an analogous connection between the
pseudo-codewords of a general LDPC code (with respect to a given
parity-check matrix), and the monomials of a certain type occurring in
the power series expansion of the edge zeta function of the associated
Tanner graph.

In summary, we believe that the here-begun study of codes from the
perspective of their efficient representation, as reflected in the
fundamental polytope, holds the key to a thorough understanding of
high performance codes and message-passing decoding algorithms.

The remainder of this paper is organized as follows. In
Section~\ref{section:LDPC}, we give background on LDPC codes and
pseudo-codewords.  Section~\ref{section:liftings} provides a technical
yet crucial result about graph covers and their associated matrices. A
characterization of pseudo-codewords in the general case via the
fundamental cone is given in Section
~\ref{section:fundamental:cone}. In Section~\ref{section:cycle:codes}
we restrict our attention to the special case of cycle codes and draw
the connection to Hashimoto's edge zeta function. We return to the
general case in Section~\ref{section:bit-even}, where we show that
every LDPC code can be realized as a punctured subcode of a code of
the type considered in the previous section. Using the results of
Section~\ref{section:cycle:codes}, we then characterize the
pseudo-codewords in the general case.

\section{Low-Density Parity-Check Codes}\label{section:LDPC}

We begin with a definition.

\begin{definition} Any subspace $C$ of $\F_2^n$ is called a {\em
    binary linear code} of {\em length} $n$.  If $C$ is described as
  the null space of some matrix $H$, i.e.,
$$
C= \left\{\vc \in \F_2^n \, \big| \, H \vc^\tr = \vzero\right\},
$$ 
  then $H$ is called a {\em parity-check matrix} for $C$.  If $H$ is
  sparse\footnote{The term ``sparse'' is necessarily vague, but
  typically one assumes that the number of 1's in each column is much
  smaller than the number of rows.  When considering a family of LDPC
  codes defined by a family $\{H_i\}_{i\geq 0}$ of $r_i \times n_i$
  matrices with $n_i$ growing increasingly large but $r_i/n_i$
  remaining fixed, ``sparse'' means that the number of 1's in the
  columns of the $H_i$ is bounded by some constant.}, we call $C$ a
  {\em low-density parity-check (LDPC)} code.
\end{definition}

Notice that the columns of $H$ correspond to the coordinates, i.e.,
{\em bits}, of the codewords of $C$, and the rows of $H$ give
relations, i.e., {\em checks}, that these coordinates must satisfy.
Although every code has many parity-check matrices, we will always fix
a parity-check matrix $H$ for each code we discuss.

The iterative decoding algorithms mentioned in
Section~\ref{section:overview} operate on a bipartite graph, called 
the {\em Tanner graph}, associated to the matrix $H$.

\begin{definition} \label{def:graph}
  An {\em undirected graph} $G = (V,E)$ consists of a set $V$ of {\em
  vertices} and a collection $E$ of 2-subsets of $V$ called {\em
  edges}.  We say $G$ has {\em multiple edges} if some 2-subset
  $\{v,w\}$ of $V$ appears in $E$ at least twice. We say two vertices
  $v, w \in V$ are {\em adjacent} if the set $\{v,w\}$ is an edge.  In
  this case, we say the edge $\{v,w\}$ is {\em incident} to both $v$
  and $w$. For $v \in V$, we write $\del(v)$ for the {\em
  neighborhood} of $v$, i.e., the collection of vertices of $G$ which
  are adjacent to $v$.  A {\em bipartite graph} with {\em partitions}
  $A$ and $B$ is an undirected graph $G = (V,E)$ such that $V$ can be
  written as a disjoint union $V = A \cup B$ with no two vertices in
  $A$ (resp., $B$) adjacent.
\end{definition}

We make the following conventions: Unless otherwise specified, our
graphs will always be undirected and our bipartite graphs will never
have multiple edges. 

\begin{definition}\label{def:LDPC}  Let $C \subseteq \F_2^n$ be the
  LDPC code determined by the (sparse) $r \times n$ matrix $H =
  (h_{ji})$. The {\em Tanner graph} $T(H)$ is the bipartite graph
  defined as follows.  The vertex set consists of the {\em bit nodes}
  $X = \{x_1, \dots, x_n\}$ and the {\em check nodes} $F = \{f_1,
  \dots, f_r\}$.  The set $\{x_i,f_j\}$ is an edge if and only if
  $h_{ji}=1$.
\end{definition}

Notice that the bit nodes in the Tanner graph correspond to the
columns of $H$, the check nodes correspond to the rows of $H$, and the
edges record which bits are involved in which checks.  In other words,
the graph $T(H)$ records the matrix $H$, and hence the code $C$,
graphically: a binary assignment of the bit nodes $(c_1, \dots, c_n)$
is a codeword in $C$ if and only if the binary sum of the values at
the neighbors of each check node is zero.  Because we have fixed a
parity-check matrix $H$ for $C$ from the start, we will also refer to
$T = T(H)$ as the Tanner graph of the code $C$.

\begin{example}\label{example:dumbbell:code}
  Let $C$ be the binary linear code of length $7$ with parity-check matrix
  \begin{equation*}
    H=\begin{pmatrix}
           1 & 1 & 0 & 0 & 0 & 0 & 0 \\
           0 & 1 & 1 & 1 & 0 & 0 & 0 \\
           1 & 0 & 1 & 0 & 0 & 0 & 0 \\
           0 & 0 & 0 & 1 & 1 & 0 & 1 \\
           0 & 0 & 0 & 0 & 1 & 1 & 0 \\
           0 & 0 & 0 & 0 & 0 & 1 & 1
         \end{pmatrix}.
  \end{equation*}

\begin{figure}
  \includegraphics[width=3in]{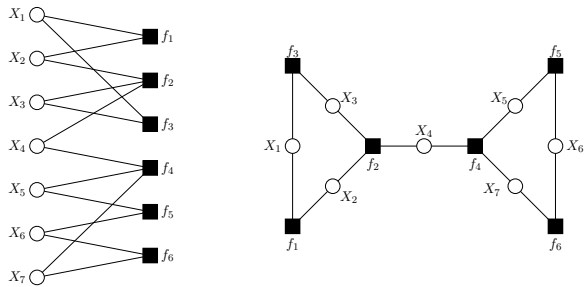}
  \caption{Two representations of the Tanner graph for the code $C$ of
  Example~\ref{example:dumbbell:code}.}
  \label{fig:dumbbell:code}
\end{figure}

  Two representations of the Tanner graph $T = T(H)$ associated to $H$
  are given in Figure~\ref{fig:dumbbell:code}, where bit and check
  nodes are represented by empty circles and filled squares,
  respectively.  The graph on the left is a traditional rendering of a
  bipartite graph, but the one on the right is easier to work with.
  The vector
  $$
  \vc = (c_1, c_2, c_3, c_4, c_5, c_6, c_7) := (1,1,1,0,0,0,0)
  $$ 
  is a codeword in $C$.  This can be checked either by computing
  $H\vc^\tr$ or by assigning the value $c_i$ to each bit node $x_i$ in
  $T$ and verifying that the binary sum of the values at the neighbors
  of each check node $f_j$ in $T$ is zero. \eexample
\end{example}

Any iterative message-passing decoding algorithm, roughly speaking,
operates as follows; see \cite{Ks-F-L} for a more precise
description. A received binary word gives an assignment of $0$ or $1$
together with a reliability value at each of the bit nodes on the
Tanner graph. Each bit node then broadcasts this bit assignment and
reliability value to its neighboring check nodes. Next, each check
node makes new estimates based on what it has received from the bit
nodes and sends these estimates back to its neighboring bit nodes. By
iterating this procedure, one expects a codeword to emerge quickly.
Notice that the algorithm is acting {\em locally}, i.e., at any stage
of the algorithm, the decision made at each vertex is based on
information coming only from the neighbors of this vertex.  It is this
property of the algorithm which causes both its greatest strength
(speed) and its greatest weakness (non-optimality).  In order to
quantify this weakness, we will need another definition.

\begin{definition}\label{def:cover} An {\em unramified, finite cover},
  or, simply, a {\em cover} of a graph $G = (V,E)$ is a graph $\tG =
  (\tV, \tE)$ along with a surjection $\pi: \tV \to V$ which is a
  graph homomorphism (i.e., $\pi$ takes adjacent vertices of $\tG$ to
  adjacent vertices of $G$) such that for each vertex $v \in V$ and
  each $\tv \in \pi^{-1}(v)$, the neighborhood $\del(\tv)$ of $\tv$ is
  mapped bijectively to $\del(v)$. A cover is called an {\em
  $M$-cover}, where $M$ is a positive integer, if
  $\left|\pi^{-1}(v)\right| = M$ for every vertex $v$ in $V$.
\end{definition}

\begin{example}\label{example:dumbbell:code:cover}

\begin{figure}
  \includegraphics[width=3in]{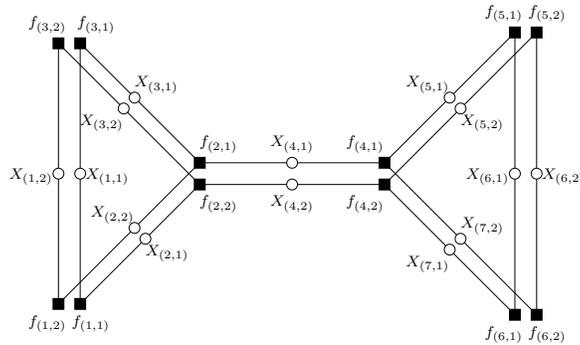}
  \caption{A $2$-cover of the code $C$ from
  Example~\ref{example:dumbbell:code}, as described in
  Example~\ref{example:dumbbell:code:cover}.} 
  \label{fig:dumbbell:code:cover}
\end{figure}

  We return to the code $C$ with chosen parity-check matrix $H$ of
  Example~\ref{example:dumbbell:code}; the corresponding Tanner graph
  $T = T(H)$ was given in Figure~\ref{fig:dumbbell:code}.  An example
  of a 2-cover (or {\em double-cover}) $\tT$ of $T$ is given in
  Figure~\ref{fig:dumbbell:code:cover}. The bipartite graph $\tT$ is
  the Tanner graph for a code $\tC$ of length $14 = 2 \cdot 7$.

  The parity-check matrix $\tH$ for the code $\tC$ associated to $\tT$
  is the $12 \times 14$ matrix 
  \begin{equation*}
  \tH = \begin{pmatrix}
        I&I&\vzero&\vzero&\vzero&\vzero&\vzero\\
        \vzero&J&I&I&\vzero&\vzero&\vzero\\
        I&\vzero&I&\vzero&\vzero&\vzero&\vzero\\
        \vzero&\vzero&\vzero&I&I&\vzero&J\\
        \vzero&\vzero&\vzero&\vzero&I&I&\vzero\\
        \vzero&\vzero&\vzero&\vzero&\vzero&I&I
        \end{pmatrix},
  \end{equation*}
  where 
  $$
  I = \begin{pmatrix}1&0\\0&1 \end{pmatrix}, \text{\quad} J =
  \begin{pmatrix}0&1\\1&0\end{pmatrix}, \text{\quad} \vzero =
  \begin{pmatrix}0&0\\0&0\end{pmatrix},
  $$
  the rows are ordered to correspond to the check nodes
  $f_{(1,1)}$, $f_{(1,2)}$, \dots, $f_{(6,1)}$, $f_{(6,2)}$, and the
  columns are ordered to correspond to the bit nodes $x_{(1,1)}$,
  $x_{(1,2)}$, \dots, $x_{(7,1)}$, $x_{(7,2)}$. \eexample
  \end{example}

Suppose $T$ is a Tanner graph for the binary linear code $C \subseteq
\F_2^n$ and $\tT$ is an $M$-cover of $T$ for some $M \geq 1$.  Let
$\tC\subseteq \F_2^{nM}$ be the binary linear code determined by $\tT$.
To indicate that the coordinates of $\F_2^{nM}$ are ordered as in
Example~\ref{example:dumbbell:code:cover} with each successive block
of $M$ coordinates lying above a single coordinate of $\F_2^n$, we
will write an element $\vx$ of $\F_2^{nM}$ as
$$
\vx = (x_{(1,1)}{:}\dots{:}x_{(1,M)}, \dots,
x_{(n,1)}{:}\dots{:}x_{(n,M)}).
$$
Every codeword in $C$ yields a codeword in $\tC$ by ``lifting'':
if $\vc = (c_1, \dots, c_n)$ is in $C$, then the vector
$$
\hvc = (c_{(1,1)}{:}\dots{:}c_{(1,M)},\dots,
c_{(n,1)}{:}\dots{:}c_{(n,M)}),
$$ 
where $c_{(i,k)} = c_i$ for $1 \leq i \leq n$ and $1 \leq k \leq M$,
is in $\tC$.  However, there are also codewords in $\tC$ which are {\em not}
liftings of codewords in $C$.  

\begin{example}\label{example:dumbbell:code:cover:codewords}
  Once again, let $C$ be the code of
  Examples~\ref{example:dumbbell:code} and
  \ref{example:dumbbell:code:cover}, and let $\tC$ be the code
  corresponding to the double-cover $\tT$ of the Tanner graph $T$ for
  $C$, as in Example~\ref{example:dumbbell:code:cover}.  The codeword
  $\vc = (1,1,1,0,0,0,0)$ of $C$ lifts to the codeword $\hvc = (1{:}1,
  1{:}1, 1{:}1, 0{:}0, 0{:}0, 0{:}0, 0{:}0)$ of $\tC$.  Although it is
  easily checked that the vector
    $$
    \tva := (1{:}0, 1{:}0, 1{:}0, 1{:}1, 1{:}0, 1{:}0, 1{:}0)
    $$ 
  is a codeword in $\tC$, it is certainly not a lifting of any
  codeword in $C$.  \eexample
\end{example}

Notice that, in general, if 
$$
\tva = (a_{(1,1)}{:}\dots{:}a_{(1,M)},\dots,
a_{(n,1)}{:}\dots{:}a_{(n,M)})
$$ 
is a codeword in the code corresponding to some $M$-cover $\tT$ of
$T$, then for any permutations $\sigma_1$, \dots, $\sigma_n$ on $\{1,
\dots, M\}$, there is an $M$-cover $\tT'$ of $T$ such that
$$
\tva' = (a_{(1,\sigma_1(1))}{:}\dots{:}a_{(1,\sigma_1(M))},\dots,
a_{(n,\sigma_n(1))}{:}\dots{:}a_{(n,\sigma_n(M))})
$$ 
is a codeword in the code corresponding to $\tT'$.  This motivates the
next definition.

\begin{definition}\label{def:pseudo-codeword}
Let $C\subseteq \F_2^n$ be a binary linear code with Tanner graph $T$
and let 
$$ 
\tva = (a_{(1,1)}{:}\dots{:}a_{(1,M)},\dots,
a_{(n,1)}{:}\dots{:}a_{(n,M)})
$$ 
be a codeword in the code $\tC$ corresponding to some $M$-cover
$\tT$ of $T$.  The {\em unscaled pseudo-codeword} corresponding to
$\tva$ is the vector $\vp(\tva) := (p_1, \dots, p_n)$ where, for $1
\leq i \leq n$, $p_i$ is the number of nonzero $a_{(i,k)}$, $1 \leq k
\leq M$.  The {\em normalized pseudo-codeword} corresponding to $\tva$
is the vector $\vomega(\tva) = (\omega_1, \dots, \omega_n)$ where each
$\omega_i$ is a rational number, $0 \leq \omega_i \leq 1$, given by
$\omega_i := \frac{1}{M}p_i$ for $1 \leq i \leq M$.
\end{definition}

\begin{example}\label{example:dumbbell:code:pseudocodewords} The
  unscaled pseudo-codeword corresponding to the codeword $\tva$ on the
  $2$-cover of Example~\ref{example:dumbbell:code:cover:codewords} is
  $(1,1,1,2,1,1,1)$.  The corresponding normalized pseudo-codeword is
  $(\frac12, \frac12, \frac12, 1, \frac12, \frac12,
  \frac12)$. \eexample
\end{example}

Notice that if $\vc$ is a codeword in our original code and $\hvc$ is
the lifting of this codeword to the code corresponding to some finite
cover of the Tanner graph, then $\vomega(\hvc) = \vc$.  Indeed, the
entries of a normalized pseudo-codeword will be entirely 0's and 1's
if and only if it comes from the lifting of some actual codeword.
Otherwise, there will be at least one entry which is non-integral.

The key issue with graph covers is that locally, any cover of a graph
looks exactly like the original graph. Thus, the fact that the
iterative message-passing decoding algorithm is operating locally on
the Tanner graph $T = T(H)$ means that the algorithm cannot
distinguish between the code defined by $T$ and any of the codes
defined by finite covers of $T$. This implies that all the codewords
in all the covers are competing to be the best explanation of the
received vector.  

To make this more precise, we assume for simplicity that we are
operating on the binary symmetric channel; the situation for other
channels is similar (see \cite{Koetter:Vontobel}).  Under this
assumption, a transmitted bit is received correctly with probability
$1 - \varepsilon$ and incorrectly with probability $\varepsilon$ where
$0 \leq \varepsilon < \frac12$.  

The goal of any decoder is to find the codeword $\vc \in C \subseteq
\F_2^n$ that best explains (in some sense) the received vector $\vy
\in \F_2^n$.  For the binary symmetric channel, a {\em maximum
likelihood} decoder will find the codeword which is closest in Hamming
distance to $\vy$.  On the other hand, because the iterative decoder
of an LDPC code acts locally on the Tanner graph associated to the
code, it allows all codewords from all finite covers to compete to be
the best explanation of $\vy$.  In a sense, it automatically lifts
$\vy$ to vectors $\hvy \in \F_2^{nM}$ for every $M \geq 1$ and
searches for a codeword $\tva$ in some code $\tC \subseteq \F_2^{nM}$
corresponding to some $M$-cover of the Tanner graph, for some $M \geq
1$, such that $\frac1M$ times the Hamming distance from (the
appropriate) $\hvy$ to $\tva$ is minimal among all codewords in all
codes corresponding to all finite covers of the Tanner graph.  Note
that even if fewer than $\lfloor\frac{d-1}{2}\rfloor$ errors have
occurred (where $d = d_{\min}(C)$ is the minimum Hamming distance of
the code), there may be codewords in covers which are at least as
close, in this sense, to $\vy$ as is the unique closest codeword.

\begin{example} \label{example:dumbbell:code:decoding}
  Consider again the code $C$ from
  Examples~\ref{example:dumbbell:code},
  \ref{example:dumbbell:code:cover}, and
  \ref{example:dumbbell:code:cover:codewords}.  Assume that we are
  transmitting over a binary symmetric channel and we receive the
  vector $\vy = (1,0,1,1,0,1,0)$.

  One can check that the codeword $\vc = (1,1,1,0,0,0,0)$ satisfies
  $d(\vy,\vc) = 3$ and that the Hamming distance from $\vy$ to any
  other codeword in $C$ is larger than $3$.  Therefore, a
  maximum-likelihood decoder would output the codeword $\vc$ when
  $\vy$ is received.
    
  But the iterative message-passing decoding algorithm allows all the
  codewords in all the codes corresponding to all the finite covers to
  compete.  In particular, the vector $\tva = (1{:}0, 1{:}0, 1{:}0,
  1{:}1, 1{:}0, 1{:}0, 1{:}0)$ from
  Example~\ref{example:dumbbell:code:cover} lies in the code $\tC$
  corresponding to the double cover $\tT$ of $T$ and is hence a
  competitor.  Letting $\hvy = (1{:}1, 0{:}0, 1{:}1, 1{:}1, 0{:}0,
  1{:}1, 0{:}0)$ be the lifting of $\vy$ to $\F_2^{14}$, we see that
  $\frac12$ times the Hamming distance from $\hvy$ to $\tva$ is also
  3.  Hence $\tva$ is just as attractive to the iterative decoder as
  $\vc$ is.  The iterative decoder becomes confused. \eexample
\end{example}

The situation observed in Example~\ref{example:dumbbell:code:decoding}
happens in general: one can easily exhibit a received vector $\vy$ and
a codeword $\tva$ in an $M$-cover for some $M$ such that $\frac1M$
times the distance from $\hvy$ to $\tva$ is at most $d(\vy,\vc)$ for
any codeword $\vc$ in the original code. As mentioned above, there is
nothing special about the binary symmetric channel, and so the above
statements can easily be generalized to other channels.

Thus, in order to understand iterative decoding algorithms, it is
crucial to understand the codewords in the codes corresponding to all
finite covers of $T(H)$.  The remainder of this paper is devoted to
this task.

\section{Liftings}\label{section:liftings}

We saw in Section~\ref{section:LDPC} above that understanding finite
covers of graphs is crucial to understanding the performance of the
iterative decoding algorithm used for LDPC codes.  The main result of
this section, Theorem~\ref{theorem:general:path:lifting}, will help us
to reach this goal.  Though it is rather technical, the remainder of
the paper hinges upon it.

We first state a lemma, the proof of which follows immediately from
the definition of an $M$-cover (Definition~\ref{def:cover}).

\begin{lemma}\label{lemma:permutation}
Let $H = (h_{ji})$ be the parity-check matrix associated to the Tanner
graph $T$ and let $\tT$ be an $M$-cover of $T$.  Let $\tH =
(h_{(j,l),(i,k)})$, $1 \leq k \leq M$ and $1 \leq l \leq M$, be the
parity-check matrix associated to $\tT$.  Then for each $i$ and $j$,
there is a permutation $\sigma_{ji}$ on $\{1, \dots, M\}$ such that
$h_{(j,l),(i,k)} = 1$ if and only if $h_{ji}=1$ and
$\sigma_{ji}(l)=k$.  Conversely, choosing permutations $\sigma_{ji}$
on $\{1, \dots, M\}$ for all $i$ and $j$ uniquely and completely
determines an $M$-cover $\tT$ of $T$ and its corresponding
parity-check matrix $\tH$.
\end{lemma}

A simple interpretation of Lemma~\ref{lemma:permutation} is that if
$H$ has associated Tanner graph $T$ and $\tT$ is an $M$-cover of $T$,
then the matrix $\tH$ associated to $\tT$ can be obtained by replacing
each 0 of $H$ with an $M \times M$ matrix of 0's and each 1 of $H$
with a suitably chosen $M \times M$ permutation matrix.  We need one
more definition before we can state the main result of this section.

\begin{definition}\label{def:path:backtrackless:edge-disjoint} Let $G
  = (V,E)$ be a graph.  Fix an ordering of the edges, so that we have
  $E = \{e_1, \dots, e_n\}$.  A sequence of edges $(e_{i_1}, \dots,
  e_{i_k})$ of $G$ is a {\em path} on $G$ if the edges $e_{i_j}$ can
  be directed so that $e_{i_s}$ terminates where $e_{i_{s+1}}$ begins
  for $1 \leq s \leq k-1$.  We say the path is {\em backtrackless} if
  for no $s$ do we have $e_{i_s} = e_{i_{s+1}}$.  We say two paths are
  {\em edge-disjoint} if they do not share an edge.
\end{definition}

The next theorem is the main result of this section.  It gives
conditions under which a collection of edges, with multiplicities, on
a graph may be lifted to a union of edge-disjoint paths on some finite
cover of the graph.  It will be used in
Section~\ref{section:fundamental:cone} to show that every vector of
nonnegative integers which lies in the fundamental cone and which
reduces modulo 2 to a codeword must be a pseudo-codeword, and that
result will be used in turn in Section~\ref{section:bit-even} to
characterize pseudo-codewords in the case in which all bit nodes in
the Tanner graph have even degree.  The proof is constructive,
providing an algorithm to produce the desired paths.

\begin{theorem}\label{theorem:general:path:lifting}  Let $T = (X \cup
  F, E)$ be a bipartite graph.  Suppose that to each $e \in E$ there
  is assigned a nonnegative integer $m_e$ such that
\begin{enumerate}\renewcommand{\labelenumi}{(H.\arabic{enumi})}
\item\label{hypoth:M_x} For each $x \in X$, there is a nonnegative
  integer $M_x$ such that $m_e = M_x$ for every edge $e$ incident to
  $x$.
\item\label{hypoth:sum:M_x:even} For each $f \in F$, the sum
  $\displaystyle{\sum_{x \in \del(f)} M_x}$ is even.
\item\label{hypoth:ineq} For each $f \in F$ and each $x \in \del(f)$,
  we have $\displaystyle{\sum_{y \in \del(f)\setminus\{x\}} M_y \geq
  M_x}$.
\end{enumerate}
Then there is a finite cover $\pi:\tT := (\tX\cup\tF,\tE) \to T$ and a
union $\Delta := \Delta_1 \cup \dots \cup \Delta_p$ of backtrackless
paths $\Delta_i$ on $\tT$ such that the endpoints of each $\Delta_i$
are in $\tX$ and such that
\begin{enumerate}\renewcommand{\labelenumi}{(C.\arabic{enumi})}
\item\label{concl:tf:at:most:once} Each $\tf \in \tF$ occurs in
  $\Delta$ at most once.
\item\label{concl:edge:at:most:once} Each $\te \in \tE$ occurs in
  $\Delta$ at most once.
\item\label{concl:empty:or:full} At each $\tx \in \tX$, either all or
  none of the edges incident to $\tx$ in $\tT$ occur in $\Delta$.
\item\label{concl:pi(Delta)=Gamma} For each $e \in E$, we have
  $\left|\pi^{-1}(e) \cap \Delta\right| = m_e$.
\end{enumerate}
\end{theorem}

\begin{proof} We will refer to $X$ as the set of bit nodes of $T$ and
  to $F$ as the set of check nodes of $T$.  Let $\Gamma$ be the
  multiset of edges of $T$ which contains, for each $e \in E$, a total
  of $m_e$ copies of $e$.  For each $f \in F$, let $N_f$ be the number
  of edges in $\Gamma$ which are incident to $f$, counted with
  multiplicity.  In other words, 
  $$
  N_f = \sum_{x \in \del(f)} M_x.
  $$ 
  Set $M := \max\left(\{M_x\, | \, x \in X\} \cup \{\frac12 N_f\, |
  \, f \in F\}\right)$.  We construct an $M$-cover $\pi:\tT \to T$ and
  the desired $\Delta$ explicitly.  The vertex set of $\tT$ is
  $$
  \{x_k \, | \, x \in X \text{ and } 1 \leq k \leq M\} \cup
  \{f_l \, | \, f \in F \text{ and } 1 \leq l \leq M\}
  $$ 
  and the map $\pi:\tT \to T$ is given by $\pi(x_k) = x$, $\pi(f_l)
  = f$.  We now need to describe the edges of $\tT$ and the disjoint
  paths $\Delta_i$.  We will first describe the edges of $\tT$ which
  are involved in the $\Delta_i$'s, and then we will describe the
  remaining edges of $\tT$.  The bit nodes of $\tT$ involved in the
  $\Delta_i$'s are $\{x_k \, | \, x \in X \text{ and } 1 \leq k \leq
  M_x\}$ and the check nodes of $\tT$ involved in the $\Delta_i$'s are
  $\{f_l \, | \, f \in F \text{ and } 1 \leq l \leq \frac12 N_f\}$.

  Start by writing out, for each $x \in X$, $M_x$ copies of the list
  $\del(x)$ of neighbors of $x$; label these lists using the bit nodes
  $x_1, \dots x_{M_x}$ of $\tT$ lying above $x$ so that $L(x_1) \dots,
  L(x_{M_x})$ are the $M_x$ copies of $\del(x)$.  Notice that there is
  a 1-1 correspondence between the edges in $\Gamma$ (with
  multiplicity) and pairs $(x_k,f)$ where $f$ occurs in $L(x_k)$.
  Similarly, write out, for each $f \in F$, one copy of the list
  $\del(f)$ of neighbors of $f$, but then replace each $x$ appearing
  in the list with the bit nodes $x_1, \dots, x_{M_x}$ of $\tT$ so
  that the list has length $N_f$; call this list $L(f)$.  Again, we
  have a 1-1 correspondence between the edges in $\Gamma$ (with
  multiplicity) and the pairs $(f,x_k)$, where $x_k$ occurs in $L(f)$.
  We will construct the $\Delta_i$'s one vertex at a time.  Each time
  we add a vertex (except for the initial vertex of each $\Delta_i$),
  we are choosing an edge from $\Gamma$ and lifting it to $\tT$, and
  so we will cross one check node off a list labeled by a bit node and
  one bit node off a list labeled by a check node.  Thus the lists
  $L(x_k)$ and $L(f)$ change as the algorithm proceeds.

  We will need some terminology and notation in the construction:
  \begin{itemize}
  \item At any given point in the algorithm and for any vertex $v$,
  let the {\em current weight} of $v$ be the number of elements in
  $L(v)$. 
  \item At any given point in the algorithm and for $x \in X$ and $f
  \in F$, set $m_f(x) := \#\{i \, | \, x_i \in L(f)\}$.
  \end{itemize}
  Notice that since, as mentioned above, the lists $L(v)$ change as
  the algorithm proceeds, the current weight of a vertex and the value
  $m_f(x)$ for $x \in X$ and $f \in F$ do as well.  At the beginning,
  the current weight of $x_k$ ($x \in X$ and $1 \leq k \leq M_x$) is
  $|\del(x)|$, the current weight of $f \in F$ is $N_f$, and $m_f(x) =
  M_x$ if $f \in \del(x)$ and $0$ otherwise.  To construct the
  $\Delta_i$'s which form $\Delta$, we proceed as follows:

  \begin{enumerate}
  \item\label{firstbit} Choose a bit node of $\tT$ whose current
    weight is at least that of every other bit node of $\tT$ and take
    it to be the first vertex in a path $P$.
  \item\label{nextcheck} Suppose we have just added the bit node $x_k$
    to $P$, where $x \in X$ and $1\leq k \leq M_x$, and that $L(x_k)
    \neq \emptyset$.  Choose a check node $f \in L(x_k)$ such that the
    current weight of $f$ is at least that of any other check node in
    $L(x_k)$.  Write down $f_s$ as the next vertex of $P$, where $s$
    is the number of times (including this one) that $f$ has been used
    so far in all of $\Delta$.  Cross $f$ off $L(x_k)$ and cross $x_k$
    off $L(f)$.
  \item\label{nextbit} Suppose we have just added the bit node $x_k$
    and then the check node $f_s$ to $P$, where $x \in X$, $f \in F$,
    $1 \leq k \leq M_x$, and $1 \leq s \leq N_f$.  Let $L(f)\setminus
    x$ denote the set of vertices in $L(f)$ which are not of the form
    $x_i$ for any $i$; Claim~\ref{claim:f:minus:x:nonempty} below
    shows that $L(f) \setminus x$ is nonempty.  Let $y_l \in
    L(f)\setminus x$ be such that $m_f(y) \geq m_f(w)$ for all $w$
    such that $w_t \in L(f) \setminus x$ for some $t$ and the current
    weight of $y_l$ is at least that of any other $y_i \in L(f)$.
    Append the vertex $y_l$ to $P$.  Cross $y_l$ off $L(f)$ and $f$
    off $L(y_l)$.  If $L(y_l)$ is now empty, then $P$ is complete and
    will be one of the $\Delta_i$'s in the disjoint union $\Delta$.
    Otherwise, return to Step~(\ref{nextcheck}).
  \item\label{done} If there are nonempty lists remaining, start over
    with Step~(\ref{firstbit}) on the remaining set of vertices.
    Otherwise, $\Delta$ is the union of the $\Delta_i$'s and the
    algorithm is complete.
  \end{enumerate}

  It is now clear from the construction and
  hypothesis~(H.\ref{hypoth:M_x}) that $\Delta = \Delta_1 \cup \dots
  \cup \Delta_p$ is a union of paths satisfying
  conditions~(C.\ref{concl:tf:at:most:once}),
  (C.\ref{concl:edge:at:most:once}) and
  (C.\ref{concl:pi(Delta)=Gamma}).
  Claim~\ref{claim:f:minus:x:nonempty} below shows that each
  $\Delta_i$ is backtrackless, and
  hypothesis~(H.\ref{hypoth:sum:M_x:even}) implies that the ending
  vertices must be bit nodes since the starting vertices are.  To see
  that condition~(C.\ref{concl:empty:or:full}) holds, let $x \in X$
  and consider two cases.  If $k>M_x$, then $x_k$ is not involved in
  $\Delta$ and so no edge incident to $x_k$ occurs in $\Delta$.  If $1
  \leq k \leq M_x$, we have $\left|\del(x)\right|$ edges incident to
  $x_k$ involved in $\Delta$.  Since the degree of $x_k$ in $\tT$ is
  to be the same as the degree of $x$ in $T$, we have that all edges
  which are to be incident to $x_k$ in $\tT$ occur already in
  $\Delta$.

  All that remains is to add additional edges to $\tT$ so that
  $\pi:\tT \to T$ is an $M$-cover.  In order for $\pi:\tT \to T$ to be
  an $M$-cover of $T$, we must have, for each bit node $x$ of $T$ and
  each $k$ with $1 \leq k \leq M$,
  $$
  |\del(x_k)| = |\del(x)|
  $$
  and
  $$
  \{f \, | \, f_l \in \del(x_k) \text{ for some } l\} = \del(x).
  $$

  Let $x \in X$.  As mentioned above, these properties hold already
  for $x_k$ with $1 \leq k \leq M_x$, and we have constructed no edges
  involving the $M-M_x$ other bit nodes $x_k$ of $\tT$.  Similarly,
  for each check node $f$ with $f \in \del(x)$, we know that exactly
  $M_x$ of the vertices $f_t$ are connected by an edge to some $x_s$,
  which means that there are $M-M_x$ vertices $f_l$ which are not
  connected by an edge to any $x_s$.  We can pair up these $M-M_x$ bit
  nodes $x_k$ and these $M-M_x$ check nodes $f_l$ any way we please.
  In particular, this will not change any bit nodes already involved
  in our $\Delta$, and when we are done doing this for each $x$, we
  will have the $M$-cover $\pi:\tT \to T$ and the $\Delta$ we seek.

The proof of Theorem~\ref{theorem:general:path:lifting} will be
complete once we have proven Claim~\ref{claim:f:minus:x:nonempty}.

\begin{claim}\label{claim:f:minus:x:nonempty} Steps~(\ref{nextcheck}) and
  (\ref{nextbit}) can always be performed without introducing a
  backtrack.  In particular, the set $L(f) \setminus x$ in
  Step~(\ref{nextbit}) of the algorithm is nonempty.
\end{claim}

\begin{proof}  For each bit node $w \in X$, let $m_f^{\mathrm{s}}(w)$
  be the value of $m_f(w)$ at the {\em start} of
  Step~(\ref{nextcheck}) and let $m_f^{\mathrm{e}}(w)$ be the value of
  $m_f(w)$ at the {\em end} of Step~(\ref{nextbit}).  For each $w \in
  X$ and each $f \in F$, let $I^{\mathrm{s}}(w,f)$ denote the
  inequality
  $$
  \sum_{y \in X \setminus\{w\}} m_f^{\mathrm{s}}(y) \geq
  m_f^{\mathrm{s}}(w) 
  $$ 
  and let $I^{\mathrm{e}}(w,f)$ denote the inequality 
  $$
  \sum_{y \in X \setminus\{w\}} m_f^{\mathrm{e}}(y) \geq
  m_f^{\mathrm{e}}(w). 
  $$ 
  Notice that at the start of the algorithm, $I^{\mathrm{s}}(w,f)$
  is true for every $w$ and $f$ by hypothesis~(H.\ref{hypoth:ineq}).
  
  Suppose $I^{\mathrm{s}}(w,f)$ holds for every $w$ and $f$ and that
  we are at the start of Step~(\ref{nextcheck}), having just appended
  $x_k$ to $P$.  We will show that we can perform
  Steps~(\ref{nextcheck}) and (\ref{nextbit}) without introducing a
  backtrack, and that the inequalities $I^{\mathrm{e}}(w,f)$ will hold
  when we are done with these two steps.  This will mean that we can
  continue to perform these steps until we are forced to move on to
  Step~(\ref{done}).

  Since each check node occurs in $L(x_k)$ at most once, we know that
  $L(x_k)$ no longer contains the check node we appended to $\Delta$
  just before we appended $x_k$.  So, since $L(x_k)$ is, by
  assumption, nonempty, Step~(\ref{nextcheck}) can be performed and it
  does not introduce a backtrack; let $f$ be the check node appended
  to $\Delta$ in that step, so that $m_f^{\mathrm{e}}(x) =
  m_f^{\mathrm{s}}(x)-1$.  Since $I^{\mathrm{s}}(x,f)$ held before
  Step~(\ref{nextcheck}), we know that there is at least one $y \neq
  x$ such that $y_i \in L(f)$ for some $i$, i.e., $L(f) \setminus x$
  is nonempty.  So Step~(\ref{nextbit}) can be performed, and we have
  $m_f^{\mathrm{e}}(y) = m_f^{\mathrm{s}}(y)-1$ for the $y \in X$
  chosen in that step.  For all other bit nodes $w$, we have
  $m_f^{\mathrm{e}}(w) = m_f^{\mathrm{s}}(w)$.  We now need to show
  that the inequality $I^{\mathrm{e}}(w,f)$ holds for every $w \in X$.
  First note that $I^{\mathrm{e}}(x,f)$ is obtained from
  $I^{\mathrm{s}}(x,f)$ by subtracting 1 from each side.  Since
  $I^{\mathrm{s}}(x,f)$ held, $I^{\mathrm{e}}(x,f)$ must also.  The
  same argument shows that $I^{\mathrm{e}}(y,f)$ holds.  Further,
  $I^{\mathrm{e}}(w,f)$ holds whenever $m_f^{\mathrm{e}}(w) =
  m_f^{\mathrm{s}}(w) = 0$ since what appears on the left-hand side of
  $I^{\mathrm{e}}(w,f)$ is certainly nonnegative.  Hence we need only
  show that $I^{\mathrm{e}}(w,f)$ holds for $w \in X \setminus\{x,y\}$
  with $m_f^{\mathrm{e}}(w) = m_f^{\mathrm{s}}(w) \geq 1$.

  So suppose $m_f^{\mathrm{e}}(w) = m_f^{\mathrm{s}}(w) \geq 1$.
  Consider first the case where $m_f^{\mathrm{e}}(v) =
  m_f^{\mathrm{s}}(v) = 0$ for all $v \in X \setminus\{x,y,w\}$. Then
  the inequality $I^{\mathrm{e}}(w,f)$ says
  $$
  m_f^{\mathrm{e}}(x) + m_f^{\mathrm{e}}(y) \geq m_f^{\mathrm{e}}(w),
  \text{\qquad i.e., \qquad} m_f^{\mathrm{s}}(x) + 
  m_f^{\mathrm{s}}(y)-2 \geq m_f^{\mathrm{s}}(w).
  $$ 
  If $m_f^{\mathrm{s}}(y) = m_f^{\mathrm{s}}(w)$, then, since
  $m_f^{\mathrm{s}}(x) + m_f^{\mathrm{s}}(y) + m_f^{\mathrm{s}}(w)$ is
  even, we know that $m_f^{\mathrm{s}}(x) \geq 2$ and so
  $I^{\mathrm{e}}(w,f)$ holds.  Otherwise, we have
  $m_f^{\mathrm{s}}(y)\geq m_f^{\mathrm{s}}(w)+1$ and so, since
  $m_f^{\mathrm{s}}(x) \geq 1$, we again see that
  $I^{\mathrm{e}}(w,f)$ holds.

  Now consider the case that there is at least one bit node $v \in X
  \setminus\{x,y,w\}$ with $m_f^{\mathrm{e}}(v) = m_f^{\mathrm{s}}(v)
  \geq 1$.  Then it is enough to show that
  $$
  m_f^{\mathrm{e}}(x) + m_f^{\mathrm{e}}(y) + m_f^{\mathrm{e}}(v) \geq
  m_f^{\mathrm{e}}(w). 
  $$
  But this is the same as
  $$
  m_f^{\mathrm{s}}(x) + m_f^{\mathrm{s}}(y) + m_f^{\mathrm{s}}(v) -2
  \geq m_f^{\mathrm{s}}(w). 
  $$ 
  Since $m_f^{\mathrm{s}}(y) \geq m_f^{\mathrm{s}}(w)$ and each of
  $m_f^{\mathrm{s}}(x)$ and $m_f^{\mathrm{s}}(v)$ is at least 1, this
  latter inequality holds and so $I^{\mathrm{e}}(w,f)$ does as well.
\end{proof}

This completes the proof of Theorem~\ref{theorem:general:path:lifting}.
\end{proof}

\section{The Fundamental Cone}\label{section:fundamental:cone} The
pseudo-codewords are described for general LDPC codes by the
fundamental cone.

\begin{definition}\label{def:fundamental:cone}
  Let $H = (h_{ji})$ be an $r \times n$ matrix with $h_{ji}\in
  \{0,1\}$ for each $j$ and $i$. The {\em fundamental cone} $\cK(H)$
  of $H$ is the set of vectors $\vnu = (\nu_1, \dots, \nu_n) \in \R^n$
  such that, for all $1 \leq i \leq n$ and $1 \leq j \leq r$, we have
  \begin{equation}\label{def:fc:nonneg}
  \nu_i \geq 0
  \end{equation}
  and
  \begin{equation}\label{def:fc:allij}
  \sum_{i' \neq i} h_{ji'} \nu_{i'} \geq h_{ji}\nu_i.
  \end{equation}
\end{definition}

\begin{remark} The matrices $H$ we consider will be parity-check
  matrices of binary linear codes.  As such, we will sometimes be
  doing computations over $\F_2$ (e.g., when deciding if a vector is a
  codeword) and sometimes over $\R$ (e.g., when deciding if a vector
  is in the fundamental cone).  Although the field over which we are
  working should usually be clear from context, we will typically
  specify it explicitly to help avoid confusion.
\end{remark}

\begin{example} \label{example:dumbbell:code:fundamental:cone}
  The fundamental cone $\cK(H)$ for the parity-check matrix $H$ of the
  code $C$ from Example~\ref{example:dumbbell:code} is
  \begin{equation*}
    \cK(H)
      = \Bigg\{(\nu_1, \dots, \nu_7) \in \R^7 \,  \Bigg| \,
      \begin{matrix}  \nu_i \geq 0 \text{ for
      $1 \leq i \leq 7$ } \\ \nu_1=\nu_2=\nu_3, \text{\quad}
       \nu_5=\nu_6=\nu_7 \\ 2\nu_1\geq \nu_4, \text{\quad}2\nu_5 \geq \nu_4
       \end{matrix} 
       \Bigg\}. 
  \end{equation*}
  Notice that the unscaled pseudocodeword $(1,1,1,2,1,1,1)$ and the
  normalized pseudocodeword
  $(\frac12,\frac12,\frac12,1,\frac12,\frac12,\frac12)$ from
  Example~\ref{example:dumbbell:code:pseudocodewords} lie in $\cK(H)$.
\eexample
\end{example}

The importance of the fundamental cone is illustrated below by
Theorem~\ref{theorem:pseudo-codewords:fundamental:cone},
Corollary~\ref{cor:normalized:in:cone} and
Theorem~\ref{theorem:rays:dense:in:cone}.

\begin{theorem}\label{theorem:pseudo-codewords:fundamental:cone} Let
  $H=(h_{ji})$ be an $r \times n$ matrix with $h_{ji} \in \{0,1\}$ for
  each $j$ and $i$, $\cK = \cK(H)$ the fundamental cone of $H$, and
  $C$ the binary code with parity-check matrix $H$.  Let $\vp = (p_1,
  \dots, p_n)$ be a vector of integers.  Then the following are
  equivalent:
\begin{enumerate}
\item $\vp$ is an unscaled pseudo-codeword.
\item $\vp \in \cK$ and $H\vp^\tr = \vzero \in \F_2^r$.
\end{enumerate}
  In other words, the unscaled pseudo-codewords are precisely those
  integer vectors in the fundamental cone which reduce modulo 2 to
  codewords.
\end{theorem}

\begin{proof} Suppose that $\vp$ is an unscaled pseudo-codeword.  Then
  there is an $M$-cover $\tT$ of the Tanner graph $T$ associated to
  $H$ and a codeword
  $$
  \tvc = (c_{(1,1)}{:} \dots {:}c_{(1,M)}, \dots, \
  c_{(n,1)}{:} \dots {:}c_{(n,M)})
  $$ 
  in $\tC$, the code associated to $\tT$, such that, for each $i$,
  exactly $p_i$ of the coordinates $c_{(i,k)}$, $1 \leq k \leq M$, are
  1.  Let $\tH = (h_{(j,l),(i,k)})$, where $1 \leq j \leq r$, $1 \leq
  i \leq n$, $1 \leq k\leq M$, and $1 \leq l \leq M$, be the
  parity-check matrix of the code $\tC$ associated to $\tT$.  For each
  $i$ and $j$, let $\sigma_{ji}$ be as in
  Lemma~\ref{lemma:permutation}, so that $h_{(j,l),(i,k)}=1$ if and
  only if $h_{ji}=1$ and $k = \sigma_{ji}(l)$.  Then the equation
  $\tH\tvc^\tr = \vzero \in \F_2^{rM}$ implies that, in $\F_2$, we
  have for each $j$ and $l$,
\begin{equation}\label{temp:observation}
  0 = \sum_{i=1}^n \sum_{k=1}^M h_{(j,l),(i,k)} c_{(i,k)} 
    = \sum_{i=1}^n h_{ji}c_{(i,\sigma_{ji}(l))}.
\end{equation}
  We shall use this observation to prove that $\vp \in \cK$ and that
  $H\vp^\tr = \vzero \in \F_2^r$.

  We first show that $\vp \in \cK$.  Clearly
  inequalities~(\ref{def:fc:nonneg}) hold for $\vnu = \vp$, and we
  must show that inequalities~(\ref{def:fc:allij}) do as well. Thus,
  we must show that we have
\begin{equation}\label{temp:ineq}
  \sum_{1 \le k \le M}\sum_{i' \neq i} h_{ji'} c_{(i',k)} \geq \sum_{1
  \le k \le M}h_{ji} c_{(i,k)}
\end{equation}
  for each $i$ and $j$.  Certainly (\ref{temp:ineq}) holds if $h_{ji}
  = 0$ or if $c_{(i,k)}=0$ for all $k$. So assume $ h_{ji} = 1$ and
  not all $c_{(i,k)}$ are zero. For each $k$ with $c_{(i,k)} = 1$, set
  $l_k := \sigma_{ji}^{-1}(k)$.  Then we have by
  (\ref{temp:observation}) that the integer sum
\begin{equation*}
\sum_{i'=1}^n h_{ji'} c_{(i', \sigma_{ji'}(l_k))}
\end{equation*}
  is even.  Hence, for each $k$ with $c_{(i,k)}=1$, there is at least
  one value of $i' \neq i$ such that $h_{ji'}=
  c_{(i', \sigma_{ji'}(l_k))} = 1$. Note that as $k$
  varies, the indices $(i',\, \sigma_{ji'}(l_k))$ are
  all distinct. Thus (\ref{temp:ineq}) holds and so $\vp \in \cK$.

  To see that $H\vp^\tr= \vzero \in\F_2^r$, sum
  (\ref{temp:observation}) over $l$ to get that for each $j$, we have
\begin{equation*}
  0 = \sum_{l=1}^M \sum_{i=1}^n h_{ji}c_{(i,\sigma_{ji}(l))}
\end{equation*}
  in $\F_2$.  After interchanging the summations over $l$ and $i$, we
  may use the fact that $\sigma_{ji}$ is a permutation and substitute
  the summation variable $l$ by $k = \sigma_{ji}(l)$ to get
\begin{equation*}
  0 = \sum_{i=1}^n h_{ji}\sum_{k=1}^M c_{(i,k)} = \sum_{i=1}^n h_{ji}
  p_i
\end{equation*}
in $\F_2$, i.e., $H\vp^\tr = \vzero \in \F_2^r$.

  Conversely, suppose $\vp = (p_1, \dots, p_n) \in \cK$ and $H \vp^\tr
  = \vzero\in \F_2^r$.  Let $T$ be the Tanner graph associated to $H$,
  and label the bit nodes of $T$ as $x_1, \dots, x_n$ to correspond to
  the $n$ columns of $H$.  For $1 \leq i \leq n$, set $M_{x_i} = p_i$.
  For each edge $e$ of $T$, there is a unique $i$, $1 \leq i \leq n$,
  such that $e$ is incident to $x_i$; set $m_e := p_i$ for this value
  of $i$.  Then hypothesis~(H.\ref{hypoth:M_x}) of
  Theorem~\ref{theorem:general:path:lifting} is satisfied.  That
  hypothesis~(H.\ref{hypoth:sum:M_x:even}) is satisfied follows
  directly from the fact that $H\vp^\tr = \vzero \in \F_2^r$.  The
  fact that $\vp \in \cK$ says that hypothesis~(H.\ref{hypoth:ineq})
  holds.  Thus Theorem~\ref{theorem:general:path:lifting} applies and
  we have a finite $M$-cover $\tT$ of $T$ for some $M \geq 1$ and a
  union $\Delta := \Delta_1 \cup \dots \cup \Delta_p$ of backtrackless
  paths on $\tT$ starting and ending at bit nodes of $\tT$ and
  satisfying
  conditions~(C.\ref{concl:tf:at:most:once})--(C.\ref{concl:pi(Delta)=Gamma})
  of that theorem.  Label the bit nodes of $\tT$ as $x_{(i,k)}$ for $1
  \leq i \leq n$ and $1 \leq k \leq M$, and let
  $$
  \tvc = (c_{(1,1)}{:}\dots{:}c_{(1,M)}, \dots, c_{(n,1)}{:}c_{(n,M)})
  \in \F_2^{nM}
  $$ be the vector given by the rule $c_{(i,k)} = 1$ if and only if
  $x_{(i,k)}$ occurs in $\Delta$, i.e., if and only if $1 \leq k \leq
  p_i$.  Then
  conditions~(C.\ref{concl:tf:at:most:once})--(C.\ref{concl:pi(Delta)=Gamma})
  ensure that $\tvc$ is a codeword in the code corresponding to $\tT$.
  Finally, we see that the unscaled pseudo-codeword associated to
  $\tvc$ is precisely $\vp$.
\end{proof}

\begin{corollary}\label{cor:normalized:in:cone} Every normalized
  pseudo-codeword lies in the fundamental cone.
\end{corollary}

\begin{proof} Let $\vomega = \frac{1}{M} \vp$ be a normalized
  pseudo-codeword, where $\vp$ is an unscaled pseudo-codeword coming
  from a codeword in the code corresponding to some $M$-cover.  Then
  $\vp \in \cK$ by
  Theorem~\ref{theorem:pseudo-codewords:fundamental:cone} and so
  $\vomega \in \cK$ since $\cK$ is a cone.
\end{proof}

\begin{theorem}\label{theorem:rays:dense:in:cone} The rays through the
  pseudo-codewords are dense in the fundamental cone.  More precisely,
  let $C$ be a binary linear code with parity-check matrix $H$, Tanner
  graph $T = T(H)$ and fundamental cone $\cK = \cK(H)$, and let $\vnu
  \in \cK(H)$. Then for any $\varepsilon > 0$, there is a 
  pseudo-codeword $\vp$ such that $||\alpha \vp - \vnu|| <
  \varepsilon$ for some $\alpha > 0$.
\end{theorem}

\begin{proof}    Let $n$ be the length of $C$, so that $\vnu
  = (\nu_1, \dots, \nu_n) \in \R^n$.  Choose $\beta \in \R$
  sufficiently large so that the vector $\vp := \vp(\beta) = (p_1,
  \dots, p_n)$, where $p_i = 2\lceil \beta \nu_i\rceil$, satisfies
  $||\alpha \vp - \vnu|| < \varepsilon$ for some $\alpha >
  0$.  For example, if $n=1$ we may take $\beta=\frac{1}{\varepsilon}$
  and $\alpha = \frac{\varepsilon}{2}$.

  We claim $\vp \in \cK$.  Certainly $p_i \geq 0$ for $1 \leq i \leq
  n$, and we must show that inequalities~(\ref{def:fc:allij}) hold for
  $\vp$.  Since $\vnu \in \cK(H)$ by assumption, we know that
  inequalities~(\ref{def:fc:allij}) hold for $\vnu$.  Multiplying
  both sides by $\beta$ and taking ceilings yields, for all $i$ and
  $j$,
  $$ 
  \left\lceil{\beta \nu_i}\right\rceil \leq \left\lceil
  \sum_{i'\neq i}h_{ji'} \beta \nu_{i'} \right\rceil \leq \sum_{i'\neq
  i}\left\lceil h_{ji'} \beta \nu_{i'} \right\rceil = \sum_{i'\neq i}
  h_{ji'} \left\lceil \beta \nu_{i'} \right\rceil.
  $$
  
  Since each $p_i$ is even, we have $H\vp^\tr = \vzero \in \F_2^r$,
  and so $\vp$ is an unscaled pseudo-codeword by
  Theorem~\ref{theorem:pseudo-codewords:fundamental:cone}. 
\end{proof}

\section{Cycle Codes}\label{section:cycle:codes} 

A binary linear code $C$ defined by a parity-check matrix $H$ is
called a {\em cycle code} if all bit nodes in the associated Tanner
graph $T(H)$ have degree $2$.  The pseudo-codewords of cycle codes
were studied by the authors in \cite{Koetter:Li:Vontobel:Walker:1}.
In this section, we review the results of that paper.  In
Section~\ref{section:bit-even}, we will show that every LDPC code can
be realized as a punctured subcode of a cycle code, and use that
relationship to give a nice characterization of the pseudo-codewords
in the general case.

The pseudo-codewords of cycle codes can be described in terms of the
monomials appearing in the {\em edge zeta function} \cite{Hashimoto},
\cite{Stark:Terras} of the {\em normal graph} \cite{Forney} of the
code.  We begin with some definitions.

\begin{definition}[\cite{Forney}]\label{def:normal:graph}
  Let $C$ be a cycle code with parity check matrix $H$ and associated
  Tanner graph $T$.  Let $X$ be the set of bit nodes of $T$ and let
  $F$ be the set of check nodes of $T$.  The {\em normal graph} of $T$
  (or of $H$, or of $C$) is the graph $N = N(T) = N(H)$ with vertex
  set $F$ and edge set $\{\del(x) \, | \, x \in X\}$.
\end{definition}

\begin{example}\label{example:dumbbell:code:normal:graph}
  Since all the bit nodes of the Tanner graph of the code $C$ from
  Example~\ref{example:dumbbell:code} have degree 2, $C$ is a cycle
  code.  The normal graph $C$ is formed by simply dropping the
  bit nodes from the Tanner graph.  It is shown in
  Figure~\ref{fig:dumbbell:code:normal:graph}.  The edge $\del(x_i)$
  is labeled by $e_i$.
\begin{figure}
  \includegraphics[width=3in]{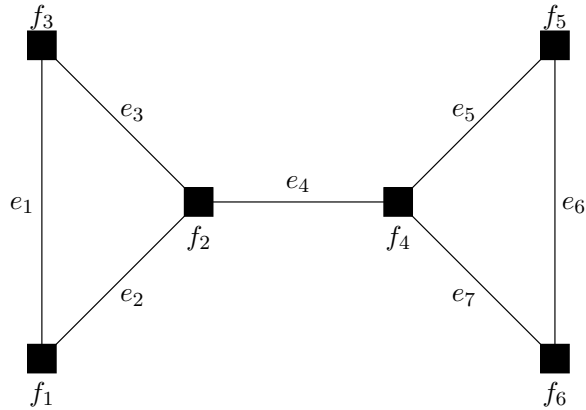}
  \caption{The normal graph of the code $C$ from
  Example~\ref{example:dumbbell:code}, as described in
  Example~\ref{example:dumbbell:code:normal:graph}.} 
  \label{fig:dumbbell:code:normal:graph}
\end{figure}
\end{example}

\begin{definition}
  Let $G = (V,E)$ be a graph.  Fix an ordering of the edges, so that
  we have $E = \{e_1, \dots, e_n\}$.  A sequence of edges $(e_{i_1},
  \dots, e_{i_k})$ of $G$ is called a {\em cycle} if the edges
  $e_{i_j}$ can be directed so that $e_{i_s}$ terminates where
  $e_{i_{s+1}}$ begins for $1 \leq s \leq k-1$ and $e_{i_k}$
  terminates where $e_{i_1}$ begins, i.e., a cycle is a path which
  starts and ends at the same vertex. We say the cycle is {\em
  edge-simple} if $e_{i_j} \neq e_{i_l}$ for $j \neq l$.  We say the
  cycle is {\em simple} if each vertex of $G$ is involved in at most
  two of the edges $e_{i_1}$, \dots, $e_{i_k}$; note that every simple
  cycle is necessarily edge-simple.  The {\em characteristic vector}
  of the edge-simple cycle $(e_{i_1}, \dots, e_{i_k})$ on $G$ is the
  binary vector of length $n$ whose $t^\text{th}$ coordinate is 1 if
  and only if $e_t$ appears as some $e_{i_j}$.
\end{definition}

The significance of the term {\em cycle code} is illustrated by the
following Lemma, which follows from Euler's Theorem
\cite[Th.~1.2.26]{West}. 

\begin{lemma}[\cite{Koetter:Li:Vontobel:Walker:1}]
  \label{lemma:cycle:code:def} {\ }
  \begin{enumerate}
  \item Let $C$ be a cycle code with Tanner graph $T$ and
    normal graph $N = N(T)$.  Then $C$ is precisely the code spanned by
    the characteristic vectors of the simple cycles in $N$.  
  \item Let $G = (V,E)$ be any graph and let $C$ be the code spanned
    by the characteristic vectors of the simple cycles in $G$.  Let $T
    = T(G)$ be the bipartite graph described as follows: The vertex
    set of $T$ is $E \cup V$.  If $e \in E$ and $v \in V$, then the
    pair $\{e,v\}$ is an edge of $T$ if and only if $e$ is incident
    to $v$ in $T$.  Then the degree in $T$ of every vertex $e
    \in E$ is 2, and $C$ is precisely the cycle code with Tanner graph
    $T$.
  \end{enumerate} 
\end{lemma}

In light of Lemma~\ref{lemma:cycle:code:def}, if $G$ is any graph, we
  call the code spanned by the characteristic vectors of the simple
  cycles in $G$ the {\em cycle code on $G$}.  In order to define the
  edge zeta function of $N$, we need some more definitions.

\begin{definition}\label{def:primitive:tailless:equivalence} 
  Let $\Gamma = (e_{i_1}, \dots, e_{i_k})$ be a cycle in a graph
  $X$. We say $\Gamma$ is {\em tailless} if $e_{i_1} \neq e_{i_k}$.
  We say $\Gamma$ is {\em primitive} if there is no cycle $\Theta$ on
  $X$ such that $\Gamma = \Theta^r$ with $r \geq 2$, i.e., such that
  $\Gamma$ is obtained by following $\Theta$ a total of $r$ times.  We
  say that the cycle $\Delta = (e_{j_1}, \dots, e_{j_k})$ is {\em
  equivalent} to $\Gamma$ if there is some integer $t$ such that
  $e_{j_s} = e_{j_{s+t}}$ for all $s$, where indices are taken modulo
  $k$.
\end{definition}

It is easy to check that any simple cycle is primitive, backtrackless
and tailless, and that the notion of equivalence given in
Definition~\ref{def:primitive:tailless:equivalence} defines an
equivalence relation on primitive, backtrackless, tailless cycles.
Also, it is clear that, up to equivalence, a cycle is backtrackless if
and only if it is tailless.  The edge zeta function of a graph is a
way to enumerate all equivalence classes of primitive, backtrackless
cycles and combinations thereof.

\begin{definition}
  \label{def:edge:zeta:function:1} \cite{Hashimoto, Stark:Terras} 
  Let $\Gamma$ be a path in a graph $X$ with edge set $E = \{e_1,
  \dots, e_n\}$; write $\Gamma = (e_{i_1}, \dots, e_{i_k})$ to indicate
  that $\Gamma$ begins with the edge $e_{i_1}$ and ends with the edge
  $e_{i_k}$. The {\em monomial of $\Gamma$} is given by $g(\Gamma) :=
  u_{i_1}\cdots u_{i_k}$, where the $u_i$'s are indeterminants. The
  {\em edge zeta function} of $X$ is defined to be the power series
  $\zeta_X(u_1, \dots, u_n) \in \Z[[u_1, \dots, u_n]]$ given by
\vspace{-1mm}
  \begin{align*}
    \zeta_X(u_1, \dots, u_n)
      &= \prod_{[\Gamma] \in A(X)}
           \big(
             1 - g(\Gamma)
           \big)^{-1},
  \end{align*}
  where $A(X)$ is the collection of equivalence classes of
  backtrackless, tailless, primitive cycles in $X$.
\end{definition}

Although the product in the definition of the edge zeta function is,
in general, infinite, the edge zeta function is a rational function
\cite{Stark:Terras}.  To make this precise, we must define the {\em
  directed edge matrix} of a graph.

\begin{definition}\label{def:directed:edge:matrix}\cite{Stark:Terras}
  Let $X = (V,E)$ be a graph with edge set $E = \{e_1, \dots, e_n\}$.
  A {\em directed graph} $\vec X$ derived from $X$ is any pair
  $(V,\vec E)$ where $\vec E = \{\vec e_1, \dots, \vec e_{2n}\}$ is a
  collection of ordered pairs of elements of $V$ such that, for $1
  \leq i \leq n$, if $e_i = \{v,w\}$ then $\{\vec e_i,\vec e_{n+i}\} =
  \{(v,w),(w,v)\}$.  (Thus we may think of $\vec X$ as having two
  directed edges, with opposite directions, for every edge of $X$.)
  The {\em directed edge matrix} of $\vec X$ is the $2n \times 2n$
  matrix $M = (m_{ij})$ with
  \begin{align*} 
    m_{ij}
      &= \begin{cases}
           1, & \text{if $\vec e_i$ feeds into $\vec e_j$ 
                      to form a backtrackless path} \\
           0, &\text{otherwise.}
         \end{cases}
  \end{align*}
  The directed edge matrix of any directed graph $\vec X$ of $X$ is
  called a {\em directed edge matrix} of $X$. 
  \end{definition}

\begin{theorem}
  \cite{Stark:Terras} The edge zeta function $\zeta_X(u_1, \dots,
  u_n)$ is a rational function. More precisely, for any directed edge
  matrix $M$ of $X$, we have
  \begin{align*}
    \zeta_X(u_1, \dots, u_n)^{-1}
      &= \det(I-UM) = \det(I-MU)
  \end{align*}
  where $I$ is the identity matrix of size $2n$ and $U =
  \mathop{\text{diag}}(u_1, \dots, u_n, u_1, \dots, u_n)$ is a
  diagonal matrix of indeterminants.
\end{theorem}

The next theorem gives the connection between the pseudo-codewords of
a cycle code and the edge zeta function of the normal graph of the
code.  Its proof was sketched in \cite{Koetter:Li:Vontobel:Walker:1},
and it is generalized in
Theorem~\ref{theorem:bit-even:pseudo-codewords} below to the case in
which all bit nodes of the Tanner graph have (arbitrary) even degree.

\begin{theorem}[\cite{Koetter:Li:Vontobel:Walker:1}]
  \label{theorem:pseudo-codewords:cycle:zeta}
  Let $C$ be a cycle code defined by a parity-check matrix $H$ having
  normal graph $N := N(H)$, let $n = n(N)$ be the number of edges of
  $N$, and let $\zeta_N := \zeta_N(u_1, \dots, u_n)$ be the edge zeta
  function of $N$. Let $p_1, \dots, p_n$ be nonnegative integers.
  Then the following are equivalent:
  \begin{enumerate}
  \item $u_1^{p_1} \dots u_n^{p_n}$ has nonzero coefficient in
    $\zeta_N$.
  \item $(p_1, \dots, p_n)$ is an unscaled pseudo-codeword for $C$
  with respect to the Tanner graph $T = T(H)$.
  \item There is a backtrackless tailless cycle in $N$ which uses the
    $i^\text{th}$ edge exactly $p_i$ times for $1 \leq i \leq N$.
  \end{enumerate}
\end{theorem}

\begin{definition}\label{def:exponent:vector}
  The {\em exponent vector} of the monomial $u_1^{p_1}\dots u_n^{p_n}$ is the
  vector $(p_1, \dots, p_n) \in \N_0^n$ of the exponents of the monomial.
\end{definition}

\begin{example} \label{example:dumbbell:code:zeta:function}
  It is shown in \cite{Koetter:Li:Vontobel:Walker:1} that the edge
  zeta function of $N$, where $N$ is the normal graph of the code $C$
  given in Example~\ref{example:dumbbell:code:normal:graph}, satisfies
  \begin{align*}
  \zeta_N&(u_1, \dots, u_7)^{-1}
    = 1 - 2 u_1 u_2 u_3 + u_1^2 u_2^2 u_3^2 - 2 u_5 u_6 u_7
         + 4 u_1 u_2 u_3 u_5 u_6 u_7 \\
    & \phantom{spa} - 2 u_1^2 u_2^2 u_3^2 u_5 u_6 u_7
                    - 4 u_1 u_2 u_3 u_4^2 u_5 u_6 u_7 + 4 u_1^2 u_2^2
    u_3^2 u_4^2 u_5 u_6 u_7 + u_5^2 u_6^2 u_7^2\\
    & \phantom{spa} - 2 u_1 u_2 u_3 u_5^2 u_6^2 u_7^2 + u_1^2 u_2^2
    u_3^2 u_5^2 u_6^2 u_7^2 + 4 u_1 u_2 u_3 u_4^2 u_5^2 u_6^2 u_7^2  -
    4 u_1^2 u_2^2 u_3^2 u_4^2 u_5^2 u_6^2 u_7^2. 
  \end{align*}
  Expanding out the Taylor series, we get the first several terms of
  $\zeta_N$:
  \begin{align*}
    \zeta_N&(u_1, \dots, u_7)
       = 1 + 2 u_1 u_2 u_3 + 3 u_1^2 u_2^2 u_3^2
           + 2 u_5 u_6 u_7 + 4 u_1 u_2 u_3 u_5 u_6 u_7 \\
      &\phantom{spa} + 6 u_1^2 u_2^2 u_3^2 u_5 u_6 u_7  
                     + 4 u_1 u_2 u_3 u_4^2 u_5 u_6 u_7
                     + 12 u_1^2 u_2^2 u_3^2 u_4^2 u_5 u_6 u_7 
                     + 3 u_5^2 u_6^2 u_7^2\\  
      &\phantom{spa} + 6 u_1 u_2 u_3 u_5^2 u_6^2 u_7^2
                     + 9 u_1^2 u_2^2 u_3^2 u_5^2 u_6^2 u_7^2 
                     + 12 u_1 u_2 u_3 u_4^2 u_5^2 u_6^2 u_7^2\\  
      &\phantom{spa} + 36 u_1^2 u_2^2 u_3^2 u_4^2 u_5^2 u_6^2 u_7^2
                     + \cdots.
  \end{align*}
  The exponent vectors of the first several monomials appearing in
  $\zeta_N$ are
  
  \begin{tabular}{l}
  (0,0,0,0,0,0,0), (1,1,1,0,0,0,0), (2,2,2,0,0,0,0), (0,0,0,0,1,1,1),
  (1,1,1,0,1,1,1), \\(2,2,2,0,1,1,1), (1,1,1,2,1,1,1), (2,2,2,2,1,1,1),
  (0,0,0,0,2,2,2), (1,1,1,0,2,2,2), \\(2,2,2,0,2,2,2), (1,1,1,2,2,2,2),
  (2,2,2,2,2,2,2), \dots. 
  \end{tabular}
  
  \noindent Note that most of these lie within the integer span of the
  codewords in $C$; for example,
  \begin{align*}
    (1,1,1,0,2,2,2)
      &= (1,1,1,0,0,0,0) + 2(0,0,0,0,1,1,1).
  \end{align*} 
  The exceptions thus far are 
  $$
  (1,1,1,2,1,1,1), (2,2,2,2,1,1,1), (1,1,1,2,2,2,2), (2,2,2,2,2,2,2).
  $$ 
  The first of these exceptions is exactly the unscaled
  pseudo-codeword of the codeword $\tva = (1{:}0, 1{:}0, 1{:}0, 1{:}1,
  1{:}0, 1{:}0, 1{:}0)$ on the double-cover $\tT$ of the Tanner graph
  $T$ in Example~\ref{example:dumbbell:code:cover}, and the rest lie
  within the integer span of this pseudo-codeword along with the
  codewords.  \eexample
\end{example}

The following corollary gives an algebraic description of the
fundamental cone in the cycle code case.

\begin{corollary}[\cite{Koetter:Li:Vontobel:Walker:1}] 
  \label{cor:newton:polyhedron:cycle}
  The Newton polyhedron of $\zeta_N$, i.e., the polyhedron spanned by
  the exponent vectors of the monomials appearing with nonzero
  coefficient in the Taylor series expansion of $\zeta_N$, is exactly
  the fundamental cone $\cK(H)$ of the code $C$.
\end{corollary}

\section{The General Case}\label{section:bit-even}

In Section~\ref{section:cycle:codes}, we saw that if $C$ is a cycle code
on a graph $N$, then the edge zeta function $\zeta_N$ of the graph $N$
has the property that the monomials appearing with nonzero coefficient
in the power series expansion of $\zeta_N$ correspond exactly to the
pseudo-codewords of $C$. It is a natural goal to find such a function
for more general LDPC codes.  In this section, we make some progress
toward this goal.

A Tanner graph is called {\em bit-even} if all the bit nodes in it
have even degree.  Let $H_0$ be a binary matrix and let $T_0 = T(H_0)$
be the associated Tanner graph.  If $T_0$ is not bit-even, let $H$ be
the matrix obtained from $H_0$ by duplicating each row of $H_0$.  Then
the Tanner graph $T$ corresponding to $H$ is obtained from $T_0$ by
duplicating all the check nodes and drawing an edge between a bit node
and a copy of a check node if and only if there was an edge between
the bit node and the original check node, so that $T$ is bit-even.
Certainly, $H_0$ and $H$ (i.e., $T_0$ and $T$) describe the same code.
Moreover, it is clear from Definition~\ref{def:fundamental:cone} that
they have the same fundamental cone, and hence, by
Theorem~\ref{theorem:pseudo-codewords:fundamental:cone}, the same
pseudo-codewords.  Thus, to describe the pseudo-codewords which arise
when we use $T_0$ to decode, we may equivalently describe the
pseudo-codewords which would arise from the (redundant) parity check
matrix giving rise to the Tanner graph $T$.  Our next task, therefore,
is to describe the pseudo-codewords associated to bit-even Tanner
graphs.

\begin{remark} Given a Tanner graph $T_0$, the procedure described
  above of duplicating all check nodes will always produce a bit-even
Tanner graph with the same fundamental cone (and hence the same
pseudo-codewords) as our original Tanner graph.  In some cases, it may
be possible to produce a Tanner graph with these properties by
duplicating only some of the check nodes.  This ``smaller'' Tanner
graph may be desirable in practice.
\end{remark}

We first describe the codewords of a code with bit-even Tanner graph
$T$ in terms of cycles on $T$.

\begin{proposition}\label{proposition:bit-even:codewords} Let $C$ be a
  binary linear code and let $T$ be a Tanner graph associated to $C$.
  Assume that $T$ is bit-even.  Then the codewords in $C$ correspond
  to disjoint unions of edge-simple cycles on $T$ such that at each
  bit node $x$ of $T$, either all or none of the edges incident to $x$
  occur.
\end{proposition}

\begin{proof} Let $X$ be the set of bit nodes of $T$ and let $F$ be
  the set of check nodes of $T$.  Fix a binary vector $\vc = (c_x)_{x
  \in X}$.  We know that $\vc$ is a codeword in $C$ if and only if,
  when we assign the value $c_x$ to every edge incident to the bit
  node $x$, the binary sum of the values of the edges incident to each
  check node is 0.  In other words, associate to $\vc$ the subgraph
  $T(\vc)$ which has as left vertices those $x \in X$ such that
  $c_x=1$, as right vertices those $f\in F$ which are joined by an
  edge in $T$ to at least one of these $x$, and as edges all the edges
  in $T$ between these $x$ and these $f$.  Then $\vc$ is a codeword if
  and only if the degree in $T(\vc)$ of each $f$ is even.  Since the
  degree of each $x$ is even by assumption, we see that $\vc$ is a
  codeword if and only if the degree of every vertex in $T(\vc)$ is
  even.  The result now follows immediately from Euler's Theorem
  \cite[Th.~1.2.26]{West}.
\end{proof}

Using Proposition~\ref{proposition:bit-even:codewords}, we may view a
binary linear code with bit-even Tanner graph as a punctured subcode
of a cycle code as follows: Let $C \subseteq \F_2^n$ be a binary
linear code with associated Tanner graph $T$, and assume that $T$ is
bit-even.  Let $\hC$ be the cycle code on $T$. Let $x_1$, \dots, $x_n$
be the bit nodes of $T$, and label the edges of $T$ (which correspond
to the coordinates of $\hC$) so that the edges incident to the bit
node $x_i$ are labeled $e_{(i,1)}$, \dots, $e_{(i,d_i)}$, where $d_i$
is the (even) degree of $x_i$.  Let $N = \sum_{i=1}^n d_i$ be the
number of edges in $T$ and define $\phi:\F_2^N \to \F_2^n$ by
$$ 
\phi(c_{(1,1)}{:} \dots: c_{(1,d_1)}, \dots,c_{(n,1)}{:} \dots{:}
c_{(n,d_n)}) = (c_{(1,1)}, \dots, c_{(n,1)}),
$$ 
i.e., $\phi$ picks off the first coordinate in each of the $n$
blocks corresponding to the $n$ bit nodes $x_i$.  Let $\hC'$ be the
subcode of $\hC$ consisting of codewords 
$$
(c_{(1,1)}{:} \dots{:} c_{(1,d_1)}, \dots,c_{(n,1)}{:} \dots{:}
c_{(n,d_n)}) 
$$ where $c_{(i,j)} = c_{(i,1)}$ for $1 \leq i \leq n$ and $1 \leq
j\leq d_i$.  Then the restriction of $\phi$ to $\hC'$ is an
isomorphism to $C$ by Proposition~\ref{proposition:bit-even:codewords}.
In other words, $C$ may be regarded as the code obtained by puncturing
the subcode $\hC'$ of $\hC$ on the positions $(i,j)$ with $2 \leq j
\leq d_i$, for $1 \leq i \leq n$.

Next, we describe the pseudo-codewords of a code with respect to a
bit-even Tanner graph $T$ in terms of $T$.

\begin{theorem}\label{theorem:bit-even:pseudo-codewords} Let $C$ be a
  binary linear code with associated Tanner graph $T$.  Assume that
  $T$ is bit-even.  Then the unscaled pseudo-codewords of $C$ with
  respect to $T$ correspond to disjoint unions of backtrackless
  tailless cycles on $T$ in which all edges incident to any given bit
  node occur the same number of times.
\end{theorem}

\begin{proof} We first set up some notation.  Let $H$ be the
  parity-check matrix for $C$ associated to $T$ and let $\cK = \cK(H)$
  be the fundamental cone.  Let $n$ be the length of $C$, let $x_1,
  \dots, x_n$ be the bit nodes of $T$, and assume $T$ has $r$ check
  nodes so that $H$ is an $r \times n$ matrix.
 
  Assume that $\vp = (p_1, \dots, p_n)$ is an unscaled pseudo-codeword
  of $C$ with respect to the Tanner graph $T$.  Then there is a
  codeword $\tvc$ in the code corresponding to some finite cover
  $\pi:\tT \to T$ of $T$ such that the unscaled pseudo-codeword
  associated to $\tvc$ is $\vp$.  Since $\tT$ is bit-even, we have by
  Proposition~\ref{proposition:bit-even:codewords} that $\tvc$
  corresponds to a union $\Delta$ of edge-simple cycles on $T$ such
  that at each bit node $x$ of $T$, either all or none of the edges
  incident to $x$ occur. Taking $\pi(\Delta)$, we get a union of
  backtrackless tailless cycles on $T$ in which all edges incident to
  any given bit node occur the same number of times, as desired.

  Conversely, suppose we are given a union $\Delta$ of backtrackless
  tailless cycles on $T$ in which all edges incident to any given bit
  node $x_i$ occur the same number, say $p_i$, of times.  Let $\vp =
  (p_1, \dots, p_n)$.  We know that $H\vp^\tr = \vzero \in \F_2^r$
  since $\Delta$ is a union of cycles, and we need to show that $\vp
  \in \cK$.  Certainly equations~(\ref{def:fc:nonneg}) hold for $\vnu
  = \vp$.  The expression $h_{ji}p_i$ counts how many edges in
  $\Delta$ go between the bit node $x_i$ and the check node $f_j$.
  Since each $\Delta_i$ is backtrackless and tailless, every time
  $\Delta_i$ goes from $x_i$ to $f_j$, it must continue to some $x_i'
  \neq x_i$.  This means that the number of edges in each $\Delta_i$
  which go between $x_i$ and $f_j$ is at most the number of edges
  which go between $f_j$ and all $x_{i'}$ with $i' \neq i$.  Thus
  $$
  \sum_{i'\neq i} h_{ji'} p_{i'}\geq h_{ji}p_i
  $$ 
  for each $i$ and $j$, i.e., equations~(\ref{def:fc:allij}) hold.
  Hence $\vp \in \cK$ and so, by
  Theorem~\ref{theorem:pseudo-codewords:fundamental:cone}, $\vp$ is a
  pseudo-codeword.
\end{proof}

Using Theorem~\ref{theorem:bit-even:pseudo-codewords}, we can describe
the pseudo-codewords of a binary linear code $C$ with respect to a
bit-even Tanner graph $T$ in terms of the exponent vectors of the
monomials appearing with nonzero coefficient in a certain power
series.  We saw above that $C$ is equal to $\phi(\hC')$, where $\hC'$
is a subcode of the cycle code $\hC$ on $C$, and $\phi$ is the map
which punctures on all positions $(i,j)$ with $2 \leq j \leq d_i$ for
$1 \leq i \leq n$.  We also have a map on the power series rings,
which we will again write as $\phi$:
$$
\phi:\Z[[u_{(1,1)}{:} \dots: u_{(1,d_1)}, \dots, u_{(n,1)}{:} \dots{:}
    u_{(n,d_n)}]] \to \Z[[u_1, \dots, u_n]].
$$
This map $\phi$ is induced by
$$
u_{(i,j)} \mapsto \begin{cases} u_i & \text{ if }j=1\\ 1 &
    \text{otherwise}.
\end{cases}
$$ 

Let
  $$
  \hzeta = \zeta_T(u_{(1,1)}, \dots, u_{(1,d_1)}, \dots, u_{(n,1)},
  \dots, u_{(n,d_n)})
  $$ 
be the edge zeta function of $T$, so that unscaled pseudo-codewords of
$\hC$ with respect to $T$ are precisely the exponent vectors of the
monomials appearing with nonzero coefficient in the power series
expansion of $\hzeta$ by
Theorem~\ref{theorem:pseudo-codewords:cycle:zeta}.  By
Theorem~\ref{theorem:bit-even:pseudo-codewords}, the unscaled
pseudo-codewords of $C$ with respect to $T$ are the unscaled
pseudo-codewords of $\hC$ with respect to $T$ in which all edges
incident to any given bit node of $T$ occur the same number of times.
If we let $\hzeta'$ be the power series obtained from $\hzeta$ by
picking off those terms with monomials of the form
$$
\prod_{\mathtiny{\begin{array}{cc}1 \leq i \leq n\\1\leq j \leq
      d_i\end{array}}} u_{(i,j)}^{p_{(i,j)}}
$$ 
with $p_{(i,j)} = p_{(i,1)}$ for $1 \leq j \leq d_i$, then the
unscaled pseudo-codewords of $C$ with respect to $T$ are precisely the
exponent vectors of the monomials appearing with nonzero coefficient
in the power series $\phi(\hzeta')$.

The above discussion is  summarized in the following theorem:

\begin{theorem} Let $C$ be a binary linear code with Tanner graph $T$,
  let $\hT$ be a bit-even Tanner graph obtained by duplicating some or
  all of the check nodes of $T$, and let $\hC$ be the cycle code on
  $\hT$.  Then $C$ is a punctured subcode of $\hC$.  Moreover, after
  choosing a suitable labeling of the $\displaystyle{\sum_{i=1}^n
  d_i}$ edges of $\hT$, where $d_i$ is the (even) degree of the
  $i^{\text{th}}$ bit node of $\hT$, the unscaled pseudo-codewords of
  $C$ with respect to $T$ are precisely those vectors $(p_1, \dots,
  p_n)$ of nonnegative integers such that
  $\displaystyle{\prod_{\mathtiny{\begin{array}{c}1 
  \leq i \leq n\\1 \leq j \leq d_i\end{array}}} u_{(i,j)}^{p_i}}$
  appears with nonzero coefficient in the power series expansion of
  the edge zeta function $\zeta_{\hT}$ of $\hT$.
\end{theorem}

\begin{remark}
When $C$ is a cycle code on a graph $N$, we saw in Section 5 that the
associated zeta function $\zeta_N$ is a rational function whose Taylor
series expansion records all pseudo-codewords of $C$. For a general
LDPC code $C$ with associated Tanner graph $T$, it would be very
interesting to find a rational function, arising combinatorially, such
that the monomials occurring in its Taylor series expansion are
precisely those in $\phi(\hat \zeta')$ constructed above.
\end{remark}

\bibliographystyle{amsplain}

\providecommand{\bysame}{\leavevmode\hbox to3em{\hrulefill}\thinspace}
\providecommand{\MR}{\relax\ifhmode\unskip\space\fi MR }
\providecommand{\MRhref}[2]{%
  \href{http://www.ams.org/mathscinet-getitem?mr=#1}{#2}
}
\providecommand{\href}[2]{#2}

\end{document}